# Multi-frequency solitons in commensurate-incommensurate photonic moiré lattices

Yaroslav V. Kartashov,[1,2] Fangwei Ye,[3] Vladimir V. Konotop,[4] and Lluis Torner[1,5]

[1]*ICFO-Institut de Ciencies Fotoniques, The Barcelona Institute of Science and Technology, 08860 Castelldefels (Barcelona), Spain*
[2]*Institute of Spectroscopy, Russian Academy of Sciences, Troitsk, Moscow, 108840, Russia*
[3]*School of Physics and Astronomy, Shanghai Jiao Tong University, Shanghai 200240, China*
[4]*Departamento de Física and Centro de Física Teórica e Computacional, Faculdade de Ciências, Universidade de Lisboa, Edifício C8, Campo Grande, Lisboa 1749-016, Portugal*
[5]*Universitat Politecnica de Catalunya, 08034 Barcelona, Spain*

We predict that photonic moiré patterns created by two mutually twisted periodic sublattices in quadratic nonlinear media allow the formation of parametric solitons under conditions that are strongly impacted by the geometry of the pattern. The question addressed here is how the geometry affects the joint trapping of multiple parametrically-coupled waves into a single soliton state. We show that above the localization-delocalization transition the threshold power for soliton excitation is drastically reduced relative to uniform media. Also, the geometry of the moiré pattern shifts the condition for phase-matching between the waves to the value that matches the edges of the eigenmode bands, thereby shifting the properties of all soliton families. Moreover, the phase-mismatch bandwidth for soliton generation is dramatically broadened in the moiré patterns relative to latticeless structures.

PhySH Subject Headings: Second order nonlinear optical processes; Optical solitons; Optical lattices.

A fascinating range of wave phenomena stemming from the geometrical properties of material landscapes are continuously discovered in diverse areas of physics. Moiré heterostructures arising as a result of mutually-rotated periodic structures are a salient example. In condensed matter they create a wealth of profound physical effects, which have established a new research area referred to as twistronics [1-4]. They are also important in cold atomic systems, Bose-Einstein condensates and optical metasurfaces [5-7]. Moiré patterns afford the possibility to explore phenomena caused by the purely geometrical properties of aperiodic (incommensurate) or periodic (commensurate) lattices, a condition that can be externally tuned by varying the rotation angle of the sublattices.

Recently it was shown [8,9] that photonic moiré lattices can be imprinted in photorefractive crystals by optical induction [10,11]. These patterns are fully reconfigurable in contrast to their material counterparts, therefore allowing the observation of the two-dimensional localization-delocalization transition (LDT) [8] of light. In one dimension, the effect had been observed for both light [12] and matter waves [13]. The phenomenon is due to the flattening of the spectral bands of the moiré pattern in the incommensurate phase, and it occurs for linear waves. However, photonic systems can be also nonlinear, affording the possibility to explore the interplay between geometry and nonlinearity. For example, lattice solitons [14,15] have been extensively studied in periodic lattices [11,16-18] and aperiodic quasi-crystals [19-22] in materials with cubic Kerr or saturable nonlinearity. Geometry-induced soliton formation effects in moiré patterns in such materials have been observed recently [23].

To date, soliton formation in parametric wave generation in materials that exhibit quadratic, rather than cubic nonlinearities, in general moiré patterns remains unexplored. Such processes involve waves at different frequencies that thus exhibit different linear bandgap spectra set by the transverse refractive index profile. Therefore, the geometrical properties of the moiré pattern impact directly the main control parameter of the wave interaction, namely the phase-matching between the interacting waves. Such solitons have been extensively studied in uniform media in several parametric-mixing processes (see [24-30] and references therein), as well as in discrete and continuous periodic systems [31-46].

In this Letter we address the existence and, importantly, the excitation of two-dimensional quadratic multiple-frequency solitons under conditions of near-phase-matched second-harmonic generation in moiré patterns imprinted in bulk crystals, and show how variations of the rotation angle and lattice depth of the sublattices impact the soliton properties. We study how the geometrical properties of the patterns modify the effective phase-matching between the fundamental frequency wave and the parametrically-generated second-harmonic wave, and how the flat-band features induced by the pattern affect the threshold for soliton existence, the power sharing between the fundamental and second-harmonic waves within the soliton families, and the threshold of soliton excitation in second-harmonic generation settings.

We study the propagation of continuous-wave light beams along the $z$-axis in a medium with $\chi^{(2)}$ nonlinearity and an imprinted refractive index distribution forming a bulk moiré pattern. The system is described by the parametrically-coupled paraxial nonlinear equations for the dimensionless electric fields of the fundamental frequency (FF), $\psi_1$, and second harmonic (SH), $\psi_2$, waves:

$$i\frac{\partial \psi_1}{\partial z} = -\frac{1}{2}\Delta\psi_1 - \psi_1^*\psi_2 e^{-i\beta z} - \mathcal{P}(\mathbf{r})\psi_1, \quad (1)$$
$$i\frac{\partial \psi_2}{\partial z} = -\frac{1}{4}\Delta\psi_2 - \psi_1^2 e^{+i\beta z} - 2\mathcal{P}(\mathbf{r})\psi_2.$$

Here $\Delta = \partial^2/\partial x^2 + \partial^2/\partial y^2$, $\mathbf{r}=(x,y)$ is normalized to the characteristic scale $a$, the propagation distance $z$ is scaled to the diffraction length $k_1 a^2$, $k_1 = n_1(\omega)\omega/c$ and $k_2 = n_2(2\omega)2\omega/c$ are the wavenumbers of the FF and SH waves at frequencies $\omega$ and $2\omega$, dimensionless complex amplitudes of the FF and SH waves are given by $\psi_{1,2} = (2\pi\omega^2\chi^{(2)}a^2/c^2)E_{1,2}$, where $E_{1,2}$ are the dimensional fields, $\chi^{(2)}$ is the relevant second-order susceptibility for the employed phase matching scheme, $\beta = (2k_1 - k_2)k_1 a^2$ is the normalized phase mismatch. Due to the difference of carrier frequencies, the optical potential $\mathcal{P}(\mathbf{r})$ is approximately two times stronger for the SH wave than for the FF wave. We assume that it represents a moiré pattern created by the superposition of two square sublattices $\mathcal{V}(R\mathbf{r})$ and $\mathcal{V}(\mathbf{r})$, where $R = R(\theta)$ is the operator of 2D rotation in the $(x,y)$ plane by the angle $\theta$: $\mathcal{P}(\mathbf{r}) = |p_1\mathcal{V}(R\mathbf{r}) + p_2\mathcal{V}(\mathbf{r})|^2$. The relative amplitudes of two sublattices are given by $p_{1,2}$ [the depth of the lattice created by shallow linear dielectric susceptibility modulation $\delta\chi^{(1)}$ is

$\max \mathcal{P} \sim \max(\delta\chi^{(1)})2\pi(\omega a/c)^2$ ], while the profile of each sublattice is given by $\mathcal{V}(\mathbf{r}) = \cos(\Omega x) + \cos(\Omega y)$. Here we set $\Omega = 2$. Examples of moiré lattices $\mathcal{P}(\mathbf{r})$ are presented in Fig. 1(a)-(c). When the rotation angle equals $\theta = \arctan[2mn/(m^2 - n^2)]$, $m, n \in \mathbb{N}$, it corresponds to a Pythagorean angle associated with the Pythagorean triple $(m^2 - n^2, 2mn, m^2 + n^2)$ and the lattice becomes exactly periodic, or commensurate [see Fig. 1(a) corresponding to $(m, n) = (2, 1)$ and Fig. 1(c) corresponding to $(m, n) = (3, 2)$]. For all other angles the lattice is aperiodic, or incommensurate [see Fig. 1(b)]. Tunable photonic moiré lattices can be readily imprinted in photorefractive crystals [7,23], including doped lithium niobate or potassium niobate crystals, as demonstrated recently [47,48]. Several techniques for critical (i.e., angle-tuning) and non-critical (e.g., temperature-tuning or QPM) phase-matching in such crystals are well-established.

The geometry of the moiré lattice impacts strongly the properties of the corresponding linear eigenmode spectrum. One peculiarity of Eq. (1) is that the FF and SH waves experience different optical potentials, thus the corresponding linear spectra are different as well. We therefore first omit the nonlinear terms in (1) and search for linear FF and SH modes as $\psi_{1,2} = w_{1,2} e^{ibz}$, where $b$ is the propagation constant. Form-factors $\chi_{FF} = U_1^{-1}[\iint |w_1|^4 d^2\mathbf{r}]^{1/2}$ and $\chi_{SH} = U_2^{-1}[\iint |w_2|^4 d^2\mathbf{r}]^{1/2}$, where $U_{1,2} = \iint |w_{1,2}|^2 d^2\mathbf{r}$ is the power, of linear mode with largest propagation constant (the most localized mode) are plotted in Fig. 1(d),(e) as a function of the rotation angle $\theta$ and the depth of the sublattice $p_2$ at fixed $p_1$. Large (small) form-factors correspond to well-localized (delocalized) modes. Pythagorean angles correspond to deeps in $\chi_{FF}, \chi_{SH}$, i.e., in commensurate lattices the linear modes are always delocalized Bloch waves. Transition from delocalization (nearly zero form-factors) to localization (form-factors $\sim 1$) occurs only in the incommensurate lattices, when $p_2$ exceeds a critical value that notably differs for the FF ($p_2^{cr} \sim 0.51$) and SH ($p_2^{cr} \sim 0.12$) waves. An exactly incommensurate lattice can be realized only in the entire $(x, y)$ plane. The size of a unit cell of a commensurate moiré is finite, but it tends to infinity as the twist angle approaches a non-Pythagorean one. In such limit the area of the Brillouin zone tends to zero resulting in nearly ideal flattening of the upper bands of the spectrum [illustrated in Fig. 1(h)]. The described band flattening, however, is different from the spectrum transformation in the tight-binding limit that can be achieved in periodic lattices that are deep enough [c.f. Fig. 1(f) and Fig. 1(g); note, that in Fig. 1(g) the FF top band appears just below its LDT transition, while the SH wave is located well-above its LDT transition]. In the commensurate tight-binding limit no localization is possible.

To calculate the linear spectrum of an incommensurate lattice, illustrated in Fig. 1(h), we used its approximation [8] by a commensurate structure generated by a Pythagorean twist angle, approaching the non-Pythagorean one. Above the LDT threshold the top band of this approximation of incommensurate lattice (and a certain number of bands below it) becomes nearly flat [already at $p_2 = 0.45$ the width $\sim 0.003$ of the FF top band in Fig. 1(h) is much smaller than FF band width $\sim 0.159$ in Fig. 1(g), while at $p_2 = 0.7$ these band widths decrease down to $\sim 10^{-9}$ and $\sim 0.041$, respectively]. In the subsequent analysis, we consider moiré patterns in which the SH wave is always above its LDT threshold, while the FF wave may be below or above its LDT threshold.

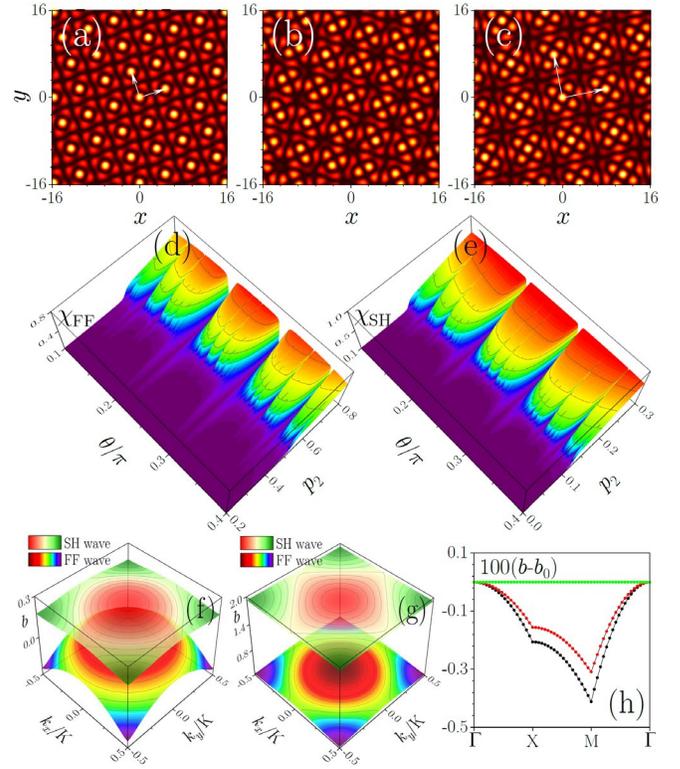

Fig. 1. Examples of moiré lattices $\mathcal{P}$ with $p_2 = 0.45$ corresponding to the rotation angles $\theta = \arctan(4/3)$ (a), $\theta = \arctan(3^{-1/2})$ (b), and $\theta = \arctan(12/5)$. White arrows show primitive translation vectors $\mathbf{e}_1 = (\pi/\Omega)[(m+n)\mathbf{i} + (m-n)\mathbf{j}]$, $\mathbf{e}_1 = (\pi/\Omega)[(n-m)\mathbf{i} + (m+n)\mathbf{j}]$ of the moiré lattice. Form-factors for the FF (d) and SH (e) linear modes supported by the moiré lattices, versus $p_2$ and $\theta$. Linear band structures for FF and SH waves in commensurate moiré lattices with $\theta = \arctan(4/3)$ at $p_2 = 0.1$ (f) and $p_2 = 0.45$ (g). Only the top bands for both waves are shown. $k_{x,y}/K$ are normalized Bloch momenta, $K = \Omega[2/(m^2 + n^2)]^{1/2}$. (h) Band structures for the FF wave in an incommensurate lattice with $\theta = \arctan(3^{-1/2})$ at $p_2 = 0.1$ (black dots), $p_2 = 0.45$ (red dots), and $p_2 = 0.7$ (green dots). To illustrate the band flattening for each $p_2$ we subtracted from $b$ the propagation constant $b_0$ corresponding to the top of the band. In all cases $p_1 = 0.3$.

To elucidate the impact of such band transformations on soliton formation, we search for solutions of Eq. (1) in the form $\psi_1 = w_1 e^{ibz}$, $\psi_2 = w_2 e^{i(\beta + 2b)z}$, where the propagation constants are chosen to describe coherent near-phase-matched interaction between the FF and the SH waves. The dependencies of soliton power $U = U_1 + U_2$ on $b$ are shown in Fig. 2(a) under conditions where the FF wave in the linear regime would be below the LDT threshold ($p_2 < p_2^{cr}$), and in Fig. 2(b) for the case when the linear FF wave would be above the LDT threshold ($p_2 > p_2^{cr}$), both for commensurate and incommensurate lattices. They reveal the existence of critical phase mismatch $\beta = \beta_{cr} > 0$ at which, within the accuracy of the numerical calculations, the threshold power for soliton formation vanishes (in uniform media it vanishes at $\beta = 0$). For solitons whose FF and SH are both originating from the semi-infinite forbidden gap, the critical phase mismatch is given by $\beta_{cr} = b_2^{upp} - 2b_1^{upp}$, where $b_1^{upp}$ and $b_2^{upp}$ are the upper edges of the top band in the linear spectra of FF and SH waves, respectively [see Fig. 1(f)-(h)]. There exists a cutoff on propagation constant for soliton existence given by $b_{co} = (b_2^{upp} - \beta)/2$ at $\beta < \beta_{cr}$ and $b_{co} = b_1^{upp}$ at $\beta > \beta_{cr}$.

When the propagation constant $b$ of the soliton approaches the cutoff value and the phase mismatch is $\beta = \beta_{cr}$, the total phase accumulation rates $d\phi_{1,2}/dz$ (where $\phi_{1,2}$ are the instantaneous phases) for the FF wave (evolving $\sim e^{ibz}$) and the SH wave (evolving $\sim e^{i(\beta+2b)z}$) simultaneously reach the upper edges $b_1^{upp}$ and $b_2^{upp}$ of the top bands in their respective spectra. The value $b = b_{co}$ depends on whether the moiré lattice is commensurate or incommensurate, and whether $p_2$ is above or below the LDT threshold. When the moiré lattice is commensurate, both waves strongly broaden, turning into the respective Bloch waves in the linear limit [an example of this behavior, at $\beta > \beta_{cr}$, is shown in Fig. 3(a)]. When the moiré lattice is incommensurate, above the LDT threshold in $p_2$ both waves turn into linear modes and remain localized [see Fig. 3(b)], while below the LDT threshold for the FF wave at least one of them broadens. If $\beta \neq \beta_{cr}$ the top edge of the band at $b \to b_{co}$ is reached by only one of the waves (for $\beta < \beta_{cr}$ this occurs for the SH wave, while for $\beta > \beta_{cr}$ this occurs for the FF wave).

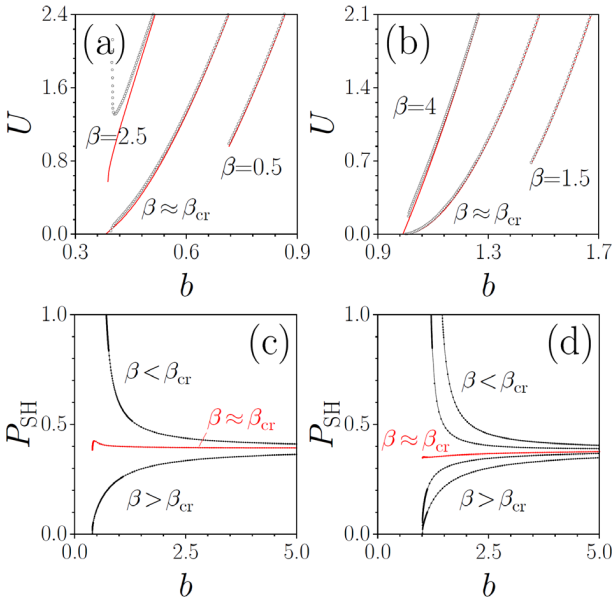

Fig. 2. Total soliton power $U$ versus propagation constant $b$ for (a) $p_1 = 0.3$, $p_2 = 0.45$ (below the LDT for the FF wave), and (b) $p_1 = 0.3$, $p_2 = 0.7$ (above the LDT for the FF wave) and different $\beta$ values. Red curves correspond to the incommensurate angle $\theta = \arctan(3^{-1/2})$, open circles correspond to the commensurate angle $\theta = \arctan(4/3)$. $\beta_{cr} \approx 1.2$ in (a) and $\beta_{cr} \approx 2.42$ in (b). Fraction of power $P_{SH}$ carried by the SH wave versus $b$ for (c) $p_1 = 0.3$, $p_2 = 0.45$, and (d) $p_1 = 0.3$, $p_2 = 0.7$ and commensurate angle $\theta = \arctan(4/3)$. The curves in (c) correspond to $\beta = 2.5$, $1.21$, $0.5$ from bottom to top. The curves in (d) correspond to $\beta = 3.5$, $2.75$, $2.45$, $2$, $1.5$ from bottom to top. We truncated the curves at $\beta \approx \beta_{cr}$ in (c),(d) very close to cutoff because their behavior in that region depends critically on the value of $\beta$, which is only known numerically (i.e., with limited accuracy).

The above result is accompanied by a qualitative change of the power sharing between the FF and the SH waves in the soliton. Figures 2(c),(d) show the fraction of power carried by the SH wave $P_{SH} = U_2/U$ as a function of $b$ for a commensurate lattice (the dependencies are qualitatively similar in the incommensurate case). At $\beta < \beta_{cr}$ the SH dominates near the cutoff, while at $\beta > \beta_{cr}$ the FF dominates. A remarkable result is that near $\beta = \beta_{cr}$ the ratio $P_{SH}$ is approximately constant for all propagation constants (for the particular value of $b$ chosen in the plot, it varies a bit only very near cutoff, but we warn that such variation is due to the unavoidable limited accuracy in the determination of $\beta_{cr}$), resembling the rigorous self-similarity rule that occurs at exact phase-matching $\beta = 0$ in the latticeless case [30,49]. Therefore, the moiré lattice, through the bandgap spectrum that it imposes, changes the conditions of the phase matching, shifting phase-matching to the value $\beta = \beta_{cr}$, where the threshold power for soliton existence vanishes. The difference in the $U(b)$ curves for commensurate and incommensurate lattices is most pronounced close to cutoff, when solutions extend across the lattice and experience the details of the refractive index landscape (its periodicity or aperiodicity), especially at $\beta > \beta_{cr}$ and below the LDT threshold. Far from the cutoff, solutions shrink to the central lattice spot in both periodic and aperiodic lattices [Fig. 3(c)]. As concerns the stability of the obtained solutions, they are stable when $dU/db > 0$. As visible in the plot, such positive slope is fulfilled by the families depicted in Fig. 2(a),(b), except for one small branch in panel (a).

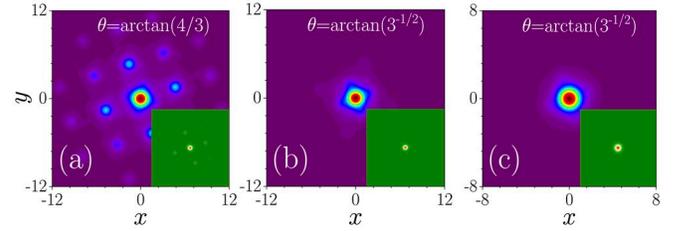

Fig. 3. Profiles of the FF and SH (insets) parts of the soliton supported by the commensurate moiré lattice with $\theta = \arctan(4/3)$ at $b = 1.011$ (a), and the incommensurate moiré lattice with $\theta = \arctan(3^{-1/2})$ at $b = 1.011$ (b) and $b = 1.5$ ($U = 4.418$) (c). In all cases $p_1 = 0.3$, $p_2 = 0.7$, $\beta = 4 > \beta_{cr}$.

Band flattening induced by the geometry of the moiré lattice dramatically impacts the behavior of the power threshold for soliton existence as illustrated in Fig. 4 comparing $U_{th}(\beta)$ curves below [Fig. 4(a)] and above [Fig. 4(b)] the LDT for commensurate and incommensurate moiré lattices. For direct comparison, with blue dashed lines we show the existence cone within which solitons exist in uniform media. One can see that moiré lattices above the LDT drastically reduce the threshold power for soliton existence, especially at $\beta > \beta_{cr}$ (thus, at $\beta - \beta_{cr} \approx 2.6$ and $p_2 = 0.7$ maximal over all angles power threshold in the lattice $\max_\theta U_{th} \approx 0.28$ is much smaller than threshold $U_{th} \approx 15.2$ in a uniform medium; the same holds true for $\beta - \beta_{cr} < 0$), and also that they greatly expand the phase-mismatch bandwidth where solitons exist, and hence can be excited, at a given power level. At fixed $p_2$ the $\beta_{cr}$ values are close for commensurate and incommensurate lattices and are not distinguishable on the scale of Fig. 4(a),(b). While at $\beta < \beta_{cr}$ the power thresholds are nearly identical in commensurate and incommensurate structures, they differ strongly at $\beta > \beta_{cr}$. This occurs even below the LDT threshold in $p_2$ [Fig. 4(a)], but is most pronounced above it [Fig. 4(b)]. Indeed, the commensurate lattice supports only delocalized linear modes, so the FF wave strongly expands near cutoff, while the total power always remains above the minimal threshold value, similarly to 2D lattice solitons in Kerr media [15]. In contrast, in the incommensurate lattice, above the LDT soliton solutions transform into linear localized modes at cutoff and therefore their power vanishes. As a result, the threshold power varies with the angle [see Fig. 4(c) for the $U_{th}(\theta)$ dependence below the LDT threshold and Fig. 4(d) for this dependence above the LDT threshold].

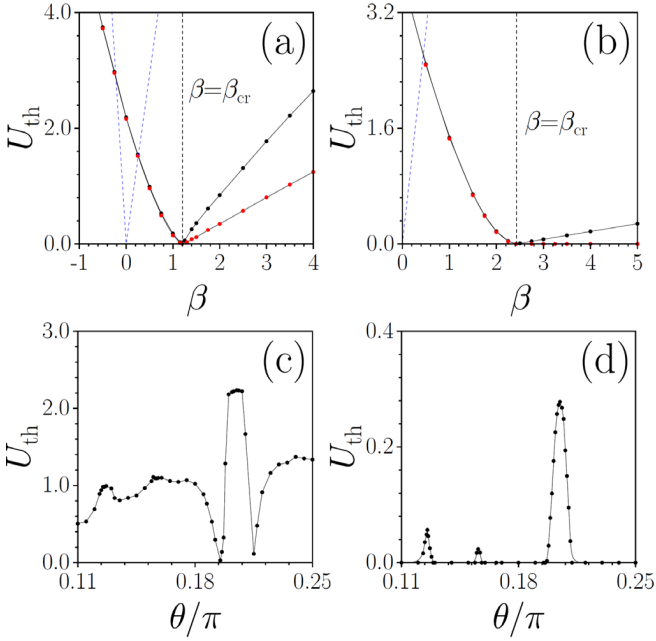

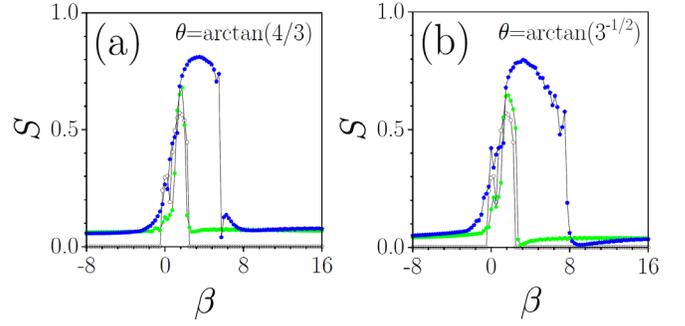

Fig. 5. Soliton content $S$ versus $\beta$ in commensurate (a) and incommensurate (b) moiré lattices below the LDT threshold for the FF wave at $p_1 = 0.3$, $p_2 = 0.45$, for input FF power $U|_{z=0} = \pi/2$ (green circles) and $U|_{z=0} = 2\pi$ (blue pentagons). For comparison, the plots also show the $S(\beta)$ dependence for uniform media at $z \to \infty$ and $U|_{z=0} = 10\pi$ (open circles).

Fig. 4. Threshold power for soliton existence $U_{\text{th}}$ versus phase mismatch $\beta$ in commensurate [black dots, $\theta = \arctan(4/3)$] and incommensurate [red dots, $\theta = \arctan(3^{-1/2})$] moiré lattices below the LDT threshold for the FF wave at $p_1 = 0.3$, $p_2 = 0.45$ (a) and above the LDT threshold at $p_1 = 0.3$, $p_2 = 0.7$ (b). The blue dashed lines in (a), (b) depict the existence cone for solitons in uniform media. $U_{\text{th}}$ versus rotation angle $\theta$ below the LDT threshold for the FF wave at $p_1 = 0.3$, $p_2 = 0.45$, $\beta = 3.5 > \beta_{\text{cr}}$ [note that the minima are not zeros] (c) and above the LDT threshold at $p_1 = 0.3$, $p_2 = 0.7$, $\beta = 5 > \beta_{\text{cr}}$ (d). At values of $\beta$ well below $\beta_{\text{cr}}$ the threshold power rapidly increases with increase of $\beta_{\text{cr}} - \beta$ and is nearly independent of the angle $\theta$.

As parametric solitons are typically excited experimentally in second-harmonic generation schemes, i.e., using FF inputs only, it is important to elucidate the excitation dynamics. Thus, we solved Eq. (1) with the input conditions $\psi_1|_{z=0} = a_1 e^{-r^2}$, $\psi_2|_{z=0} = 0$. Figure 5 compares approximate soliton content $S$, defined as a ratio of the power $U$ contained in the eventually excited parametric soliton within a ring of radius $r_S = 2\pi/\Omega$ at sufficiently large distance and the input power $U|_{z=0}$ (carried only by the FF wave). The dependence $S(\beta)$ illustrates the obtained phase-mismatch bandwidth corresponding to the dynamical soliton excitation. For direct comparison we also show the $S(\beta)$ dependence in a uniform medium (open circles) [49]. One can see that even below the LDT threshold (i.e., at $p_2 < p_2^{\text{cr}}$ for the FF wave), in the moiré lattice the bandwidth at $U|_{z=0} = \pi/2$ is comparable with that obtained in a uniform medium at $U|_{z=0} = 10\pi$. In other words, the power required for similarly efficient soliton excitation is reduced by more than order of magnitude in moiré structure. The phase-mismatch bandwidth expands with $U|_{z=0}$ and always remains larger in the incommensurate structure in comparison with the commensurate one [compare Fig. 5(b) with 5(a)]. As expected, above the LDT threshold the bandwidth expands even further because the diffraction for the FF wave is drastically reduced.

In conclusion, the geometrically-induced localization properties of moiré patterns impact strongly the existence and excitation conditions of quadratic multi-frequency solitons. Under suitable conditions the threshold for soliton excitation is drastically reduced, and the corresponding phase-mismatch bandwidth is largely enhanced. Remarkably, this is the case even in moiré patterns in which only one of the interacting waves propagates under conditions above its LDT. Here we addressed second-harmonic generation configurations, but we anticipate that the uncovered insight is relevant for all self-trapping mechanisms mediated by parametric wave-mixing processes.